Title: ERROR PROPAGATION IN EXTENDED CHAOTIC SYSTEMS
--------------------------------------------------------------------------------
\documentstyle[preprint,aps]{revtex}

\newcommand {\be}{\begin{equation}}
\newcommand {\ee}{\end{equation}}
\newcommand {\vL}{v_{\rm L}}
\newcommand {\vF}{v_{\rm F}}

\begin{document}
\draft

\title{Error Propagation in Extended Chaotic Systems}
\author{
Alessandro Torcini, Peter Grassberger\\
{\it Theoretische Physik,
Bergische Universit\"at-Gesamthochschule Wuppertal,
\\ D-42097 Wuppertal, Germany}\\}
\author{
Antonio Politi\\
{\it Istituto Nazionale di Ottica and INFN,\\ I-50125 Firenze, Italy}\\}

\date{\today}

\maketitle
\begin{abstract}
A strong analogy is found between the evolution of localized disturbances in
extended chaotic systems and the propagation of fronts separating different
phases. A condition for the evolution to be controlled by nonlinear
mechanisms is derived on the basis of this relationship. An approximate
expression for the nonlinear velocity is also determined by extending
the concept of Lyapunov exponent to growth rate of finite perturbations.
\end{abstract}

\vskip 1 truecm


In the last years the study of front propagation in spatially
extended systems has known a renewed interest, due to the
relevance of spreading fronts for the emergence of spatial
structures (patterns) in non-equilibrium systems \cite{cross}.
In particular, simple reaction-diffusion models seem to be
appropriate for describing propagation phenomena in different
fields, such as fluid dynamics, liquid crystals
\cite{crystals}, epidemics \cite{mollison}, chemical reactions,
crystal growth \cite{benjacob} and biological aggregation
\cite{levine}.
Several mathematical models which describe the spreading of a
disturbance into unstable (or metastable) steady states have been
studied in detail, in order to uncover the mechanisms underlying
the propagation of fronts \cite{piskunov,aronson,saarl,benguria}.

The main result of these studies can be summerized with reference to
the one dimensional equation
\be
u_t = u_{xx} + g(u)
\label{dif_eq}
\ee
where $g(u) \in C^1[0,1]$, $g(0)=g(1)=0$. If $g > 0$ in $(0,1)$,
then $u=0$ is an unstable fixed point, while $u=1$ is a stable one.
In this case, any sufficiently localized initial perturbation $u(x,t=0)$
generates a propagating front joining
the unstable to the stable state (Fig.1). A linear stability analysis
shows that the front can have any speed $\vF$ larger than a minimal
value $\vL$ which depends on the behavior of $g(u)$ at $u=0$. However,
very often velocities larger than $\vL$ require special initial
conditions to be realized, so that the ``physical" speed is exactly
$\vF=\vL$.
In the following we shall call $\vL$ the {\it linear velocity}. Whether
it is selected or not depends on the behavior of $g(u)$ for $u>0$.
In particular, it has been shown in \cite{collet} that convexity of
$g(u)$ is sufficient for $\vF=\vL$. Intuitively, we can say that, if
$g'(0) >g'(u)$ for all $u>0$, the front is ``pulled" by the initial
growth of $u$ and, otherwise, it is ``pushed" by the faster growth of
finite $u$ \cite{paquette}.

In the present Letter we consider a different problem, namely the
propagation of some perturbation in a chaotic system (Fig. 2). Thus the front
does not separate two different phases, since the system is chaotic
(and hence unstable) on both sides of the front. More precisely, we
consider two realizations of a 1-d coupled map lattice (CML)
\cite{cml} which differ only locally in the initial conditions, and we watch
the spreading of the relative deviation. In spite of the obvious difference
with the situation discussed above, we will show that there are surprising
similarities.
In particular, the derivative $g'(u=0)$ will be replaced by the Lyapunov
exponent. In order to formulate (heuristically) a condition equivalent to
$g'(0) >g'(u)$, we will introduce a new indicator of the
sensitivity to finite perturbations. We shall see that there exists again
a minimal velocity $\vL$, and that the ``physical" velocity $\vF$ can be
larger that $\vL$ only if this indicator grows with the perturbation.

The CML is written as
\be
x_i^{n+1} = f({\tilde x}_i^n)
\label{cml1}
\ee
\be
{\tilde x}_i^n = (1-\varepsilon) x_i^n + \frac{\varepsilon}{2}
(x_{i-1}^n + x_{i+1}^n)
\label{cml2}
\ee
where $i$ and $n$ indicate the discrete space and time variables,
and $\varepsilon$ the diffusive coupling parameter. We use periodic boundary
conditions on a chain of length $L$, $x^n_{i}=x^n_{i\pm L}$. The function
$f(x)$ is assumed to be a map of some interval into itself. We have chosen this
for numerical convenience. We are confident that the basic features
can be generalized to continuous systems. Disturbances spreading
with $ v_{\rm F} \ge v_{\rm L}$ in CMLs have been observed for the first
time in \cite{pol1}. Here, instead of considering the map $G(x)$ studied in
\cite{pol1}, we shall discuss two simpler examples:
\noindent
the ``generalized Bernoulli shift"
\be
    f(x)= r x \quad {\rm mod}\; 1 \quad
\label{bern}
\ee
and the circle map
\be
    f(x)= (x + \alpha) \quad {\rm mod}\; 1 \quad .
\label{circle}
\ee
The propagation of infinitesimal disturbances is governed by
the evolution in tangent space,
\be
    u^{n+1}_i = f^{\prime}( {\tilde x}_i^n)
    \left[ (1-\varepsilon) u^n_i + {\varepsilon \over 2}
    \left( u^n_{i+1} + u^n_{i-1} \right) \right]
                  \label{lincml}
\ee
where $f^\prime = df/dx$. Instead of considering an initially localized
perturbation, we shall first refer to a perturbation decaying exponentially
for $i\to\infty$,
\be
   u_i^0 \sim e^{-\mu i} \;.
\ee
Its temporal growth depends on $\mu$,
\be
   u_i^n \sim e^{\lambda(\mu) n - \mu i} \;.
\ee
The position of the front is defined as the rightmost site where $u_i^n$
is larger than some arbitrarily fixed constant ${\cal O}(1)$.
This gives for its velocity
\be
   V(\mu) = {di\over dn}= \frac{\lambda(\mu)}{\mu}\;. \label{vmu}
\ee
For an absolutely unstable system we have $\lambda(\mu=0)>0$ ,
\footnote{Throughout this letter,
all Lyapunov exponents are maximal ones, and all perturbations are
assumed to be typical so that they grow with maximal rate. There exist of
course also atypical perturbations the growth of which is governed by
non-leading Lyapunov exponents \protect{\cite{lepri}}, but they will be
neglected.} so that
$V(\mu)$ diverges for $\mu\to 0$. This is intuitively obvious:
an almost flat front will appear to move with arbitrarily large velocity.
On the other hand, it can be shown \cite{lepri} that, for nearest-neighbour
coupling, $V(\mu)\to 1$ for $\mu\to\infty$.

We now want to determine the speed $\vL$ when the initial perturbation
is localized near $i=0$ and still infinitesimal (the case of finite
perturbations will be discussed later).
Since we expect that any front will have a leading edge where it is
infinitesimal and exponentially decaying with some exponent $\mu_0$,
we have $\vL = V(\mu_0)$.

To determine $\mu_0$ and $\vL$, we need the (maximal) {\it co-moving}
Lyapunov exponent $\Lambda(v)$ \cite{deissler}. For a given $v$, this gives
the local growth rate of a disturbance in a reference
frame moving with velocity $v$, $u_i^n \sim e^{\Lambda(v) n}$ if $i=vn$.
The selected front speed is such that a disturbance neither
grows nor decreases at $\vL$, i.e. $\Lambda(\vL)=0$. In order to
express this in terms of $\lambda(\mu)$ and $\mu$, we recall that they
are related to $\Lambda(v)$ through the Legendre transformation
\cite{pol2,bohr}
\be
   \Lambda(v)=\lambda(\mu) - \mu \frac{d \lambda (\mu) }{d \mu} \quad ;
        \quad v =  \frac{d \lambda (\mu) }{d \mu}
   \label{legendre}
\ee
Therefore, the derivative of $V(\mu)$ is directly related to
the co-moving exponent,
\be
   \frac{d V}{d \mu} = \frac{1}{\mu} \left( \frac{d \lambda}{d \mu} -
   \frac{\lambda}{\mu} \right) = -\frac{\Lambda(v)}{\mu^2} \;.    \label{vl0}
\ee
Using $\Lambda(\vL)=0$, we now see that $dV/d\mu=0$ at a value $\mu_0$ for
which $v(\mu)=\vL$, and since
$\Lambda(v)$ is convex (being a Legendre transform), this will be the
unique minimum of $V(\mu)$. Finally, we can write
\be
   \vL = \frac{\lambda(\mu_0)}{\mu_0} = \left(
          \frac{d \lambda (\mu) }{d \mu} \right)_{\mu=\mu_0}\;.
                    \label{vl}
\ee
Thus as long as we can consider a perturbation as infinitesimal, it is
the lowest possible speed which is selected, which justifies us
calling it the ``linear velocity''.

This expression for $\vL$ is identical to that found in Ref.
\cite{saarl} for the propagation into unstable steady states, provided
that $\lambda(\mu)$ and $\mu$ are identified with the complex part of
the frequency and of the wavevector, respectively. Thus, the relation
$\lambda=\lambda(\mu)$ plays essentially the role of a dispersion relation
\cite{lepri,bohr}.

Recalling that for closed systems, $\Lambda(v)$ is always a
decreasing function (limiting us to $v \ge 0$ for symmetry
reasons) and that $\Lambda(v=0) = \lambda(0)$ \cite{deissler},
we can readily deduce from Eq.~(\ref{vl}) that $\vL$ is
defined if and only if the system is absolutely unstable, i.e. $\lambda(0)
> 0$. As can be seen from Fig.~3, $V(\mu)$ steadily increases with
$\mu$ and $V(\mu \to 0) \to - \infty$ if the local dynamics is
not chaotic ($\lambda(0)<0$). A negative velocity indicates that the
perturbation regresses instead of propagating: the system is absolutely
stable.

Finally, we consider localized and {\it finite} initial perturbations.
We call the corresponding front velocity $\vF$. Since any front will
have an infinitesimal leading edge, we have to expect that $\vF =
V(\mu^*)$ for some value $\mu^*$. It is hard to see how $\mu^*$ could
be smaller than $\mu_0$, whence we just have to distinguish two
possibilities: the ``linear" (or ``pulled") case with $\mu^*=\mu^0$
and $\vF=\vL$, and the ``nonlinear" (or ``pushed") case with $\mu^*>\mu^0$,
$\vF>\vL$.

In order to see which case is realized in a particular model, we simulate
two chaotic configurations $\{ x^n_i \}$ and $\{ y^n_i \}$
initially differing in a limited region of the chain (typically
50 sites in chains of $\geq 1024$ sites) and coinciding elsewhere.
The front position after $n$ iterations is defined as
\be
R(n) = \max \{ i : |x^n_i - y_i^n | \ge \theta \}  \;.
\ee
where $\theta$ is a preassigned threshold $ << 1$.
The front velocity is then defined as
\be
v_{\rm F} = \lim_{n \to \infty} { {R(n)} \over n}
\label{vF}
\ee
We have verified that $\vF$ is independent of the amplitude of the
initial perturbation $\delta_0$ and of the value of the threshold $\theta$ when
they are varied from $10^{-14}$ to $10^{-1}$.

In this way we measured $\vF$ and $\vL$ for several
CML models and couplings. As expected, we found always
$\vF \ge \vL$. In most cases, $\vF = \vL$ (this was found for
logistic, cubic and tent coupled maps for all tested values of the
parameters and of $\varepsilon$), but
we have also identified a class of maps (namely, models (\ref{bern}),
(\ref{circle}) and the map $G$ studied in Ref.\cite{pol1}) where the strict
inequality $v_{\rm F} > v_{\rm L}$ is found to hold. The common characteristic
of these maps is that $f^\prime(x)$ exhibits a narrow peak (or even a
$\delta$-singularity). Moreover, in system (\ref{bern}) \cite{torc} and in
Ref. \cite{pol1} a transition between the two above regimes is
found upon varying a parameter of the map. For map (\ref{circle}), such a
transition cannot occur since $v_{\rm L}$ is always zero, the map being
marginally stable.  However, also in this case we can observe
a finite $v_{\rm F}$ for a range of $\alpha$ and $\varepsilon$ values.
This fact stresses even more that this propagation mechanism is not
related to local chaoticity, i.e. to sensitive dependence on local
and infinitesimal perturbations. The unpredictability resulting from
the spreading of perturbations does not result here from local production of
entropy but from entropy transport.

In order to determine when the nonlinear mechanism is likely to prevail
against the linear one, we reconsider a heuristic conjecture of van
Saarlos \cite{saarl} for fronts propagating into unstable steady and
homogeneous states. He observed that $\vF>\vL$ only if the local growth
rate of small but finite perturbations increases with their amplitude.

In our case the linear local growth rate of perturbations is
represented in the limit of small coupling $\varepsilon$ by the Lyapunov
exponent of the single map $\lambda_0$, which can be defined as
\be
\lambda_0 = \lim_{\delta \to 0} \left< \log \left|
\frac{f(x+\delta/2) - f(x-\delta/2)}{\delta} \right| \right>
=  < \log | f^\prime(x) | >
\label{lyap}
\ee
where $< \dots >$ is the average over the invariant measure of the map.
If we are interested in the evolution of finite disturbances $\Delta$
the average growth-rate will be given by
\be
I(\Delta) = \left< \log \left|
\frac{f(x+\Delta/2) - f(x-\Delta/2)}{\Delta} \right| \right>
=  < A(x,\Delta) > \;.
\label{indic}
\ee
Obviously, $\lim_{\Delta \to 0} I(\Delta) = \lambda_0$.
Let us first consider map (\ref{bern}). There, we have
$$
A(x,\Delta) = \cases {
 \log \left[ {(1-r \Delta) \over \Delta} \right] \quad ,
& if $x \in [1/r - \Delta/2, 1/r +\Delta/2] \equiv C(\Delta,r)$ \cr
\lambda_0 = \log (r) \quad , & otherwise\cr}
$$
Therefore, the indicator $I$ is given by
\be
I(\Delta) =
\int_{x \in C}  {\rm d}x \enskip \nu (x) \enskip
\log \left[ {(1-r \Delta) \over \Delta} \right]
+ \int_{x \not\in C}  {\rm d}x \enskip \nu (x) \enskip
\log (r)
\ee
where $\nu (x)$ is the invariant measure.

The expression is more compact for the circle map, because there
$\lambda_0=0$ and the invariant measure is flat, so that
\be
 I(\Delta) = \Delta \log \left[ {(1-\Delta) \over \Delta} \right] \;.
\label{indcircle}
\ee
This is positive for $0 < \Delta < 1/2$, and is an increasing
function at small $\Delta$. Therefore at small but finite $\Delta$
we have a positive growth-rate in spite of the stability against
infinitesimal perturbations. This is always the case when we
consider maps like (\ref{bern}), (\ref{circle}) or $G$. Conversely,
for all the other maps we looked at (i.e., logistic and tent maps), we found
$I(\Delta) < \lambda_0$ for all finite $\Delta$-values (see Fig. 4).
Accordingly, we can conjecture that whenever a nonlinear propagation
mechanism has been observed, the quantity $I(\Delta)$ is an increasing
function at small $\Delta$. If, instead, $I(\Delta) < \lambda_0$
for any $\Delta$, propagation in the correspondig CML will take
place with velocity $v_{\rm F} = v_{\rm L}$ for any coupling costant
$\varepsilon$.
Nonlinear propagation of perturbations can arise only if finite
disturbances are, in average, amplified faster than infinitesimal
ones, i.e. by a factor $>\exp[\lambda_0]$ during a single iteration.

In order to give a quantitative estimate of $v_{_F}$ we have to
take into account the coupling between different sites.
We have seen that the linear velocity is the minimum value of
$V(\mu)$ which is obtained from the growth rate $\lambda(\mu)$.
If $\vF>\vL$, the exponential slope $\mu^*$ of the leading edge is
larger than the value $\mu_0$ where $V(\mu)$ is minimal. The
main effect of nonlinearities is to change $\lambda(\mu)$
into a function $\lambda(\mu,\Delta)$ which coincides
with it along the leading edge of the front (where $\Delta$ is
infinitesimal) but becomes different as $\Delta$ becomes large.
Our main assumption now is that we have just to replace $\lambda(\mu)$ with
a suitable average over $\lambda(\mu,\Delta)$. The average has to be taken
over the $\Delta$ range where $I(\Delta)>\lambda(0)$ and which thus
``pushes" the front.

The main problem in this assumption is of course that $\Delta$
is a fluctuating quantity. In order to make it practically
applicable, we have to resort to a mean field approximation.

By assuming that the perturbation decays exponentially as
\be
\Delta_i^n = {\rm e}^{- \mu i} \Phi_i^n \; ,
\label{cmldist}
\ee
from Eqs.(2) and (3), we find that it evolves in time according to
\be
\Delta_i^{n+1} =
      |f({\tilde x}_i^n + {1\over 2} {\tilde \Delta}_i^n) -
       f({\tilde x}_i^n - {1\over 2} {\tilde \Delta}_i^n)|=
      {\tilde  \Delta}_i^n  \;
         {\rm e}^{A({\tilde x}_i^n,{\tilde  \Delta}_i^n)}
\label{fcml}
\ee
where
\be
  {\tilde  \Delta}_i^n = {\rm e}^{- \mu i}
    \left((1-\varepsilon) \Phi_i^n+{\varepsilon\over 2}
     ( \Phi_{i-1}^n\,e^\mu + \Phi_{i+1}^n\,e^{-\mu}) \right) \;.
\ee
We now introduce a mean field approximation by assuming that $\Phi_i^n$ is
independent of $i$, and $A(x,\Delta)$ equal to its average over $x$.
This allows us to rewrite Eq. (\ref{fcml}) as
\be
   \Phi^{n+1}= \Phi^{n} \left[ (1-\varepsilon) +
     \varepsilon \; {\rm cosh}(\mu) \right]\; {\rm e}^{I(\Delta)}
\ee
Performing an average over the range $D$ of $\Delta$ where $I(\Delta)>0$,
we obtain an effective Lyapunov exponent
\be
   \lambda_c (\mu) = \log \left[ (1-\varepsilon) +
   \varepsilon {\rm cosh}(\mu) \right] + \frac{1}{|D|}
   \int_D {\rm d} \Delta \; I(\Delta)\;,\quad |D|= \int_D {\rm d} \Delta
               \label{nllyap}
\ee
and, in analogy with the linear case,
\be
V_c(\mu) = \frac{\lambda_c (\mu) }{\mu}  \; .
\label{vnlmu}
\ee
Just like $V(\mu)$, $V_c(\mu)$ is a convex function
with a unique minimum. It is thus natural to assume
that the selected velocity for the front will be given by
the minimum of (\ref{vnlmu})
\be
v_{_F} = {\rm min}_\mu V_c (\mu)  \; .
\label{vnl}
\ee
The value $\mu_c$ where $V_c(\mu)$ is minimal would be equal to $\mu^*$ if
Eq. (\ref{cmldist}) would hold with the same $\mu$ in the leading edge and in
the
pushing region. This, however, need not be the case and we indeed found
$\mu_c < \mu^*$ in general.

In Fig. 5, the numerical results are reported toghether with
the predictions obtained from Eq. (\ref{vnl}) for the circle
map with two different values of $\alpha$.
The agreement between simulation and theoretical results
is reasonably good for large coupling $\varepsilon$. However,
it can be seen that the front propagates only for $\varepsilon$
larger than a certain threshold $\varepsilon_c(\alpha)$.
Equation (\ref{vnl}) does not predict such a transition which
can be attributed to the particular structure
of the invariant measure for model (\ref{circle}) for
$\varepsilon < \varepsilon_c(\alpha)$. The invariant measure becomes
extremely irregular below threshold and this does
not allow any more a "synchronization" of the motion of
the disturbances, as necessary to observe a front
propagation. Obviously, this cannot be recovered from a
mean field analysis.
An analogous comparison for map (\ref{bern}) with $\varepsilon = 1/3$ is
reported in Fig.~6. The overall behaviour of the velocity provided by
Eq.~(\ref{vnl}) is in agreement with that of the measured $v_{\rm F}$.
More precisely, the theoretical predictions are larger than $v_{\rm L}$ for
any value of the parameter $r$.

In conclusion, we have demonstrated that the propagation
of perturbations in chaotic systems is very similar to
the propagation of fronts between steady states. This
includes the possibility of ``nonlinear" selection of
velocity. We have verified that an extremely crude estimate
of the influence of nonlinearities on the velocity gives
surprisingly good agreement with simulations of several
coupled map lattices.

\acknowledgements{
We thank Dr. Stefano Lepri for useful discussions.
One of us (A.T.) gratefully acknowledges the European Economic Community
for the research fellowship No ERBCHBICT941569 "Multifractal
Analysis of Spatio-Temporal Chaos".
}

\begin{figure}
\noindent
{\bf Fig.1}: A typical front connecting an unstable with a stable region.
\end{figure}

\begin{figure}
\noindent
{\bf Fig.2}: A typical chaotic state ($x$), a perturbed state ($y$),
and their difference. The front separates the perturbed from the not yet
perturbed region.
\end{figure}

\begin{figure}
\noindent
{\bf Fig.3}: Velocities $V(\mu)$ versus $\mu$ for the coupled piecewise
linear maps (\protect{\ref{bern}}) with $\varepsilon=1/3$.
The solid curve refers to the absolutely unstable situation ($r>1$),
the dashed line to the marginally stable case ($r=1$) and
the dash-dotted one to the absolutely stable case ($r < 1$).
\end{figure}

\begin{figure}
\noindent
{\bf Fig.4}: Nonlinearity indicator $I(\Delta)$ for the single maps:
logistic map at the crisis (solid line); tent map (dashed line);
circle map (\protect{\ref{circle}}) with $\alpha=[1-(\sqrt{5}-1)/2]$
(dotted line); generalized Bernoulli shift (\protect{\ref{bern}}) with $r=1.10$
(dash-dotted line).
\end{figure}

\begin{figure}
\noindent
{\bf Fig.5}: Front velocities for circle coupled maps as a function
of the coupling parameter $\varepsilon$:
measured velocity $v_{\rm F}$ (crosses) and
theoretical prediction $v_T$ (circles). Figure (a) refers to a
map parameter $\alpha= [1-(\sqrt{5}-1)/2]$ and (b) to
$\alpha = [1-(\sqrt{5}-1)/2]/8$.
\end{figure}

\begin{figure}
\noindent
{\bf Fig.6}: As in Fig.~5 for coupled piecewise linear
map with $r > 1$ ($\varepsilon=1/3$).
In this case the linear velocity $v_{\rm L}$ is reported
too (solid line), since it is positive.
\end{figure}

\end{document}